# Inferring Causal Relationships to Improve Caching for Clients with Correlated Requests: Applications to VR

AGRIM BARI, The University of Texas at Austin, USA
GUSTAVO DE VECIANA, The University of Texas at Austin, USA
YUQI ZHOU, Purdue University, USA

Efficient edge caching reduces latency and alleviates backhaul congestion in modern networks. Traditional caching policies, such as Least Recently Used (LRU) and Least Frequently Used (LFU), perform well under specific request patterns. LRU excels in workloads with strong temporal locality, while LFU is effective when content popularity remains static. However, real-world client requests often exhibit *correlations* due to shared contexts and coordinated activities. This is particularly evident in Virtual Reality (VR) environments, where groups of clients navigate shared virtual spaces, leading to correlated content requests.

In this paper, we introduce the *grouped client request model*, a generalization of the Independent Reference Model that explicitly captures different types of request correlations. Our theoretical analysis of LRU under this model reveals that the optimal causal caching policy depends on cache size: LFU is optimal for small to moderate caches, while LRU outperforms it for larger caches. To address the limitations of existing policies, we propose *Least Following and Recently Used (LFRU)*, a novel online caching policy that dynamically infers and adapts to causal relationships in client requests to optimize evictions. LFRU prioritizes objects likely to be requested based on inferred dependencies, achieving near-optimal performance compared to the offline optimal Belady policy in structured correlation settings.

We develop VR based datasets to evaluate caching policies under realistic correlated requests. Our results show that LFRU consistently performs at least as well as LRU and LFU, outperforming LRU by up to 2.9× and LFU by up to 1.9× in certain settings. Specifically, across different cache capacities, LFRU achieves performance gains ranging from 1× to 2.9× over LRU and 0.6× to 1.9× over LFU, highlighting its adaptability and robustness across various scenarios.

## 1 INTRODUCTION

***Managing shared edge caching.*** Efficient management of shared edge caches plays a crucial role in modern networked systems, reducing latency, alleviating backhaul network congestion, and improving client experience. However, ensuring data availability for multiple clients on the limited shared edge caches with diverse and dynamic request patterns presents a significant challenge. To maximize the overall performance, there are two decisions to be made, first, about what data to

Authors' addresses: Agrim Bari, The University of Texas at Austin, Austin, TX, USA, agrim.bari@utexas.edu; Gustavo de Veciana, The University of Texas at Austin, Austin, TX, USA, deveciana@utexas.edu; Yuqi Zhou, Purdue University, West Lafayette, IN, USA, zhou1168@purdue.edu.

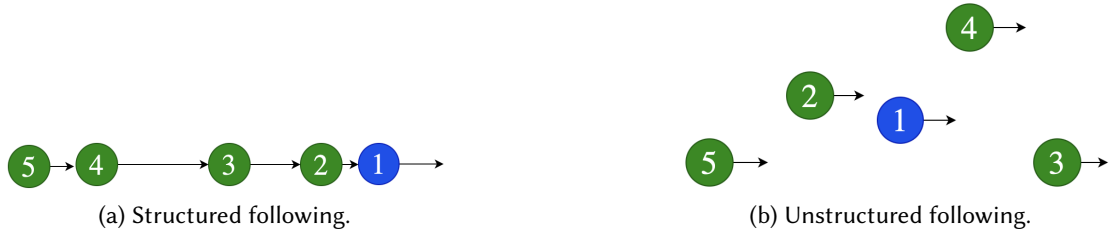

Fig. 1. Different types of correlations in a VR environment, Client 1 is the leader client and remaining clients are followers.

place in the cache (placement), second, what data to evict when storage constraints are reached (eviction).

*Existing heuristic caching policies.* Traditional caching policies rely on well-established heuristics to determine cache placement and eviction. Least Recently Used (LRU) prioritizes the eviction of objects accessed least recently, making it effective when request patterns exhibit strong temporal locality. Least Frequently Used (LFU) retains the most commonly requested objects over time, excelling in settings where demand popularity remains relatively stable. However, in many scenarios, client data requests exhibit correlations which can be dynamically changing.

*Correlation in client requests and applications.* In many real-world applications, the patterns of request of client data are not independent but exhibit strong correlations due to shared contexts, coordinated activities, or inherent behavior of the system. For example, in collaborative Virtual Reality (VR) environments, correlations in client requests can arise explicitly, such as when students follow their teacher through a VR space and request the same data objects with some delay. Alternatively, correlations may emerge implicitly when multiple clients independently visit a popular virtual location, such as New York, leading to overlapping content requests. Similarly, in collaborative editing platforms such as Google Docs or GitHub, employees working on shared documents or repositories often access overlapping data files in rapid succession.

A similar structure exists in edge computing, where client-generated tasks rely on microservices that must be loaded into the edge server's memory before execution. In many applications, clients invoke specific sequences of microservices due to inherent task dependencies. For instance, a request for an authentication service is frequently followed by requests for data processing or storage services, with multiple clients exhibiting similar access patterns. These correlations, whether arising from user behavior, system architecture, or application workflows, present an opportunity to optimize caching strategies, improving efficiency and reducing service latency.

*Different client request patterns in VR.* In VR environments, client request patterns can exhibit different forms of correlation, as illustrated in Fig. 1. One common scenario, which we denote as *structured following*, occurs when followers replicate their leader's requests sequentially. For example, in a virtual museum setting, a teacher guiding students through exhibits results in each student accessing the same content in the same order as the teacher. In contrast, *unstructured following* arises when followers are distributed around the leader, leading to a staggered and less predictable request pattern, such as when students explore a VR environment independently, but still request content influenced by the teacher's interactions.

In this work, we introduce a unified model denoted *grouped client request model* to capture structured and unstructured following behaviors. This model extends the traditional Independent Reference Model (IRM) by explicitly accounting for correlations among client requests. Furthermore, we propose a caching policy that dynamically infers causal relationships between client requests.

In scenarios characterized by structured following, our approach significantly outperforms the conventional LRU policy and others by prioritizing the retention of objects likely to be requested by (follower) clients based on these inferred dependencies.

### 1.1 Related work

***Analytical works on caching policies and approximation.*** We restrict our review to the most relevant papers in our work. [4, 13] summarize significant early work in the design of caching policies, and [10] describes analytical methods and evaluation results for the performance assessment of caching strategies. The aim of any caching policy is to achieve efficient cache utilization. This efficiency is measured primarily by the cache hit rate, which is the averaged fraction of data object requests for which the data object is in the cache when requested.

Besides hit rate, other design objectives for caching policies are ease of implementation, low operational overhead, and adaptability to fluctuations in access/request patterns. An important difference among caching policies is in what they evict when the cache is full. Under Least Recently Used (LRU), the cache is consistently updated to hold the most recently requested data objects, enabling it to leverage the temporal locality of data object requests. Notably for LRU under the Independent Reference Model (IRM), where each data object is requested independently of any past requests, the invariant distribution assuming data objects of the same size [15] and an approximation for the hit rate [3, 5, 6, 9] have been obtained. In particular, [5] describes the working-set approximation for hitting probabilities, the fraction of requests for a data object for which the object is in the cache. This approximation has been shown to be accurate as the number of objects scales [6, 9]. In this paper, we define a new working set approximation and use that to find the hit probability for requests from different clients under a grouped client request model defined later.

Under the IRM model, for a fixed cache capacity with same-size data objects, caching the most popular data objects is optimal for causal policies [1]. Least Frequently Used (LFU) performs optimally under stationary regimes of request patterns by replacing cached data objects based on the frequency measurements of past requests. An interesting work by [12] shows that a variant of LRU that infers the instantaneous request rate subject to the history of requests can come arbitrarily close to the optimal LFU algorithm. [14] shows that even for strongly correlated request patterns, LFU is still optimal among causal policies. However, while LFU may be effective in stationary scenarios where access patterns remain relatively constant, it may struggle to perform optimally in non-stationary regimes where the dynamics of data access change over time.

***Machine Learning-Based Caching Approaches.*** Recent advancements in Machine Learning (ML) have significantly influenced the development of intelligent and adaptive caching strategies. Approaches such as those in [7, 19, 20] leverage supervised learning to predict content popularity and optimize caching decisions. [19] focuses on long-term predictions, utilizing deep learning models to forecast future content demand. By contrast, [7] emphasizes the ability to adapt to short-term, immediate fluctuations in user demands, particularly in edge networks. While both methods demonstrate the potential of ML in improving caching performance, they are heavily reliant on predictive models that require substantial amounts of training data. Moreover, they typically struggle to adapt to rapidly changing workloads, especially when faced with unpredictable shifts in user behavior. This limitation is mitigated by our proposed LFRU policy, which does not depend on predictive modeling. Instead, LFRU adapts in real-time based on observed correlations between client requests, making it more responsive to dynamic workloads without the need for extensive training data.

The authors of [16] and [17] employ reinforcement learning (RL) to make cache eviction decisions based on factors such as access patterns, object size, and content popularity. These RL-based approaches dynamically adjust their caching decisions in response to changing network conditions, thereby optimizing content delivery efficiency. While RL-based methods are effective in handling dynamic environments, they typically require continuous retraining to maintain accuracy as access patterns evolve. This presents a challenge in rapidly changing contexts, such as VR applications, where request patterns can fluctuate significantly. LFRU, however, does not rely on retraining or predictive modeling. Instead, it infers causal relationships between client requests in real-time, ensuring efficient cache eviction decisions even as access patterns change unpredictably.

The works most closely aligned with our approach are [11] and [8]. In [11], the authors use a neural network to model the inter-relationships between content requests. By learning patterns of dependencies among requests, [11] aims to optimize cache eviction decisions by prioritizing content that is more likely to be requested soon. In contrast, LFRU builds similar relationships across client requests but does so in a computationally more efficient manner than neural networks. This makes LFRU a more lightweight solution, especially in resource-constrained environments.

In [8], cache decisions are made based on the Follow-The-Regularized-Leader algorithm, which optimizes the selection of actions by combining historical data with regularization penalties. Similarly, LFRU makes cache eviction decisions based on the observed behavior of clients and infers which clients are more likely to be followed by others. While both methods aim to optimize decision-making based on past behavior, LFRU offers a more direct and computationally less expensive approach by focusing on observed client correlations rather than relying on complex models or algorithms.

Finally, [18] explores hybrid strategies that integrate machine-learned predictions with traditional caching techniques. This combination seeks to improve the competitive ratio of caching algorithms despite potential inaccuracies in predictions. While hybrid methods have shown promise in improving caching performance, they still rely on predictive models, which can incur higher computational costs and may struggle with rapidly changing workloads. LFRU, in contrast, offers a purely observational approach that does not rely on predictions or complex models, providing a more efficient alternative in environments where quick adaptation and low computational overhead are critical.

## 1.2 Paper contributions and organization

In summary, we make the following key contributions:

- We introduce the *grouped client request model*, a generalization of IRM that captures different types of correlations in client requests.
- We derive a *working-set approximation* for computing hit probabilities under LRU and show that LFU is suboptimal for large caches in correlated request settings.
- We propose *Least Following and Recently Used (LFRU)*, a lightweight online caching policy that adapts to structured correlations when present, outperforming both LRU and LFU across cache sizes.
- We develop VR-based datasets to capture different types of correlated client requests and empirically show that LFRU improves cache hit ratios by up to 2.9× over LRU and 1.9× over LFU.

The remainder of this paper is organized as follows. In Section 2, we describe the system model and present the working-set approximation for LRU under the grouped client request model. This section also includes an approximation for calculating hit probabilities for different clients. In Section 3, we introduce our caching policy that leverages inferred causal relationships. We then empirically evaluate and compare different caching policies using a dataset emulating a VR environment in Section 4. Finally, we conclude the paper in Section 5.

## 2 SYSTEM MODEL, ANALYSIS AND SIMULATION RESULTS

### 2.1 Model for cache

We shall consider a simple cache with capacity $b$ bytes. The cache stores various data objects to serve future requests from a client population. Due to practical and cost constraints, the cache capacity typically is not enough to store all data objects.

### 2.2 Model for grouped client request patterns

We consider a fixed set of data objects, denoted by $\mathcal{D}$, where $D$ is the total number of data objects that a population of clients can request. Clients are represented by the set $C = \{1, 2, \ldots, C\}$, where $C$ is the total number of clients. Each client belongs to exactly one of several distinct and non-overlapping groups, denoted by $\mathcal{G} = \{1, 2, \ldots, G\}$, where $G$ is the total number of groups.

For each group $g \in \mathcal{G}$, let $\mathcal{D}^g$ represent the set of data objects that can be requested by clients in the group, and $C^g$ represent the set of clients in the group. Each group has a designated leader client, denoted by $l^g$, along with a number $f^g$ of followers.

For a data object $d \in \mathcal{D}$, we let $\mathcal{G}^d \subseteq \mathcal{G}$ denote the subset of groups that may request the data object $d$. Note that while groups do not overlap in terms of clients, they may share common data objects.

DEFINITION 1 (GROUPED CLIENT REQUEST MODEL). *Each group's leader generates requests data objects independently of other requests, following a stationary Poisson process. Followers in each group make the same data object requests but with a possibly random delay relative to the time the leader made the request.*

*Formally, let $\lambda^g(d)$ denote the arrival rate of requests for data object $d \in \mathcal{D}^g$ by the leader $l^g$. The total arrival rate of requests generated by leader $l^g$ is denoted as $\lambda^g = \sum_{d \in \mathcal{D}^g} \lambda^g(d)$. The probability that the leader requests data object $d \in \mathcal{D}^g$ is denoted as $p^g(d) = \frac{\lambda^g(d)}{\lambda^g}$.*

*We define $\Delta_i^g$ as a random variable representing the delay (or possible advancement) between the request of the $i$-th follower and the leader of group $g$ for the same data object. The joint distribution of these delays across all followers in group $g$ is $(\Delta_i^g : i = 1, \ldots, f^g)$. These delays can have an arbitrary joint distribution, independent but not identically distributed, or independent and identically distributed (i.i.d). The grouped sequence of requests from group $g$ is modeled as a Marked Poisson Point Process (MPPP):*

$$A^g = \left( (A_n^g, D_n^g, (\Delta_{i,n}^g : i = 1, \ldots, f^g)) : n \in \mathbb{Z}^+ \right) \sim MPPP(\lambda^g),$$

*where $A_n^g$ is the arrival time of the $n$-th request from the leader of group $g$, $D_n^g$ is a random variable representing the data object requested by the leader, $(\Delta_{i,n}^g : i = 1, \ldots, f^g) \sim (\Delta_i^g : i = 1, \ldots, f^g)$ represents the joint distribution of delays for the $n$-th request from the followers of group $g$, relative to when the leader made the request. These delays associated with followers requests are independent across different request instances $n$.*

*The overall sequence of arrivals across all groups is denoted as $A = (A^g : g \in \mathcal{G})$. The processes $A^g$ for different groups $g \in \mathcal{G}$ are independent MPPPs. The probability that the leader of group $g$ requests data object $d \in \mathcal{D}^g$ is $\mathbb{P}(D_n^g = d) = p^g(d)$.*

*Finally, we let $s(d)$ denote the size of data object $d$, and let $\Lambda = (\lambda^g(d) : g \in \mathcal{G}, d \in \mathcal{D}^g)$ represent the vector of request rates for each group leader and data object.*

REMARK 1 (GENERALITY OF THE GROUPED CLIENT REQUEST MODEL). *The proposed model is flexible and can represent various scenarios where client requests are grouped/correlated. For example:*

(1) *Independent Reference Model (IRM): If $f^g = 0$ for all $g \in \mathcal{G}$, the model reduces to the Independent Reference Model, where client requests are independent and uncorrelated.*

(2) *Structured follower requests: If the delays $\Delta_i^g$ are fixed and structured, such as $\Delta_i^g = \delta_i + \Delta_{i-1}^g$ for $i > 1$ where $\delta_i$ are positive constants, then each leader's request is followed by a predictable sequence of follower requests. This models scenarios like a teacher guiding students in a VR environment, where both the teacher and students request data objects that fall within their field of view. Here, the students follow the teacher in a fixed sequence, making requests one after another.*

(3) *Unstructured random follower requests: If the delays follow a random distribution, such as $\Delta_i^g \sim U[\alpha_i, \beta_i]$, where $\alpha_i$ can be negative, the model represents a scenario where follower requests are randomly distributed around the leader's request. This captures a more dynamic VR setting, where students follow a teacher but do not adhere to a strict structure. Students may request data before or after the teacher.*

*These examples demonstrate the model's ability to handle both deterministic and random correlated patterns for client group requests.*

REMARK 2 (EXTENDING THE GROUPED CLIENT REQUEST MODEL). *While this model assumes that followers always request data after their leader, it can be easily extended to allow followers to randomly opt out of following the leaders requests. This is a potential direction for our future work.*

## 2.3 Hit probabilities for leaders and followers under LRU under the grouped client request model

The Least Recently Used (LRU) policy evicts the least recently accessed data object in the cache when a new object is requested, provided that the new object is not already in the cache and there is no space available to cache. Suppose $d$ is requested at time 0. Under LRU, assuming the cache orders data objects from most recently used to least recently used, $d$ initially occupies the bottom position. Over time, after some data objects (other than $d$) are requested, $d$ moves to the top of the cache and is subsequently evicted, provided that it is not requested again before this happens. Now, we let $T_b(d)$ be a random variable representing the amount of time it would take to accumulate other unique data objects requests excluding $d$ so as to fill the cache. This variable, $T_b(d)$, is referred to as *characteristic time* of the data object $d \in \mathcal{D}$.

Quantifying $T_b(d)$ is important for determining the probability that $d$ will still be in the cache under the LRU policy. In the literature, two approximations simplify this calculation, and these approximations become more accurate as the total number of data objects, $D$, increases (i.e., when $D \gg 1$). For further details, see [3, 9].

**Approximation 1:** For $D \gg 1$, the characteristic time $T_b(d)$ concentrates around its mean value. Therefore, $T_b(d)$ can be approximated by a deterministic value, $t_b(d)$, for each data object $d$.

**Approximation 2:** The dependence of $t_b(d)$ on the specific data object $d$ can be neglected. This approximation is commonly used and justified in the literature [3, 9], particularly when the request probability $\sum_{g \in \mathcal{G}^d} p^g(d)$ is small relative to that of the remaining data objects request probabilities. This approximation becomes exact when the request probabilities are uniform.

Thus we introduce $t_b^*$ as the mean characteristic time for any data object. Given the above approximations, $t_b^*$ satisfies the following equation:

$$b = \sum_{d \in \mathcal{D}} \mathbb{P}\left(\text{data object } d \text{ was requested at least once in } [-t_b^*, 0]\right) s(d). \quad (1)$$

where the right hand side captures the size of set of data objects present in the cache at time 0 without loss of generality and since under LRU a data object stays in the cache for $t_b^*$ amount of time after it is requested, we focus on the case whether a data object was requested at least once in the time interval $[-t_b^*, 0]$. This is referred to as the working set approximation, which has been shown to be accurate as the number of data objects scales. We define an event $V_\tau^{l^g,i}$ as follows: the leader client $l^g$ of group $g$ makes a request at time $\tau$, and $i$–th follower's request for the same data object does not occur within $[-t_b^*, 0]$. This condition is equivalent to $\{\tau + \Delta_i^g \notin [-t_b^*, 0]\}$. Let $p(d, t_b^*)$ denote the probability that data object $d$ is requested at least once in $[-t_b^*, 0]$. We compute it in the following lemma.

LEMMA 1 (WORKING SET APPROXIMATION). *Consider a cache of size $b$ that follows the Least Recently Used (LRU) eviction policy and serves a grouped client request pattern (see Section 2.2 for details). The probability that a data object $d$ is requested at least once within the time interval $[-t_b^*, 0]$ is given by*

$$p(d, t_b^*) = 1 - e^{-\sum_{g \in \mathcal{G}^d} \int_{-\infty}^{\infty} \lambda_\tau^g(d, t_b^*) \, d\tau}, \tag{2}$$

*where $\lambda_\tau^g(d, t_b^*) = \lambda^g(d) q_\tau^g(t_b^*)$. The function $q_\tau^g(t_b^*)$ is defined as:*

$$q_\tau^g(t_b^*) = \begin{cases} 1 - \mathbb{P}\left(\bigcap_{i=1}^{f^g} V_\tau^{l^g,i}\right) & \text{if } \tau \in [-\infty, -t_b^*) \cup (0, \infty), \\ 1 & \text{if } \tau \in [-t_b^*, 0]. \end{cases} \tag{3}$$

*Under Approximations 1 and 2, the characteristic time $t_b^*$ is determined by solving the fixed-point equation:*

$$b = \sum_{d \in \mathcal{D}} p(d, t_b^*) s(d). \tag{4}$$

PROOF. To derive $p(d, t_b^*)$, we first compute the probability that no client in a group $g \in \mathcal{G}^d$ requests data object $d$ within the interval $[-t_b^*, 0]$.

The leader of group $g$ generates requests for $d$ as a PPP with rate $\lambda^g(d)$. Consider the leader of group $g$ requests $d$ at time $\tau$, then its $i$-th follower requests it at time $\tau + \Delta_i^g$. A request for data object $d$ from a client in group $g$ can fall in the interval $[-t_b^*, 0]$ either due to the leader's request occurring within the interval or due to a follower's request. Formally, we define $q_\tau^g(t_b^*)$ as the probability that at least one client in group $g$ requests $d$ in $[-t_b^*, 0]$, given that the leader made a request at time $\tau$. This probability is given by

$$q_\tau^g(t_b^*) = \begin{cases} 1 - \mathbb{P}\left(\bigcap_{i=1}^{f^g} V_\tau^{l^g,i}\right) & \text{if } \tau \in [-\infty, -t_b^*) \cup (0, \infty), \\ 1 & \text{if } \tau \in [-t_b^*, 0]. \end{cases} \tag{5}$$

Since the leader requests arrive according to a PPP, we can define an inhomogeneous thinned PPP that models at least one client in group $g$ requesting $d$ within $[-t_b^*, 0]$, with rate $\lambda_\tau^g(d, t_b^*) = \lambda^g(d) q_\tau^g(t_b^*)$. We can now compute the probability that no client in group $g$ requests $d$ in $[-t_b^*, 0]$ as:

$$e^{-\int_{-\infty}^{\infty} \lambda_\tau^g(d, t_b^*) \, d\tau}. \tag{6}$$

Since the requests from different groups are independent, the probability that $d$ is requested at least once in $[-t_b^*, 0]$ is

$$p(d, t_b^*) = 1 - \mathbb{P}(\text{No client in } \mathcal{G}^d \text{ requests } d \text{ in } [-t_b^*, 0]), \tag{7}$$

$$= 1 - e^{-\sum_{g \in \mathcal{G}^d} \int_{-\infty}^{\infty} \lambda_\tau^g(d, t_b^*) \, d\tau}. \tag{8}$$

Finally, under the LRU policy, the total size of cached data objects at time 0 is $\sum_{d \in \mathcal{D}} p(d, t_b^*) s(d)$. Under approximations 1 and 2, we can now find the characteristic time $t_b^*$ which satisfies

$$b = \sum_{d \in \mathcal{D}} p(d, t_b^*) s(d). \tag{9}$$

□

We now use this lemma to compute the hit probability for a data object requested by a client in a group. Let $p^L(d, g)$ denote the hit probability for data object $d$ when requested by the leader $l^g$ of group $g$. Similarly, let $p^F(d, g; i)$ represent the hit probability for a request from the $i$-th follower in group $g$.

For two clients $c_1, c_2 \in C^g$ in group $g$, we define an event $E^{c_1, c_2}$ as follows: client $c_1$ makes a request for a data object at time 0, and client $c_2$'s request for the same data object does not occur within the time interval $[-t_b^*, 0]$. For example, if $c_1$ is the leader $l^g$ and $c_2$ is the $i$-th follower in group $g$, then $E^{l^g, i}$ represents the event that the leader generates a request at time 0, but the $i$-th follower's request does not fall within $[-t_b^*, 0]$. This condition is equivalent to $\{\Delta_i^g \notin [-t_b^*, 0]\}$.

THEOREM 1 (HIT PROBABILITY FOR CLIENTS). *Consider a cache of size b that follows the Least Recently Used (LRU) eviction policy and serves request per the grouped client request model. Under some approximations, the hit probability for a data object d requested by leader $l^g$ of group g is given*

$$p^L(d, g) = 1 - \left[1 - p(d, t_b^*)\right] \left[\mathbb{P}\left(\bigcap_{i=1}^{f^g} E^{l^g, i}\right)\right].$$

*Similarly, under the same approximations, the hit probability for a data object d requested by i-th follower of group g is*

$$p^F(d, g; i) = 1 - \left[1 - p(d, t_b^*)\right] \left[\mathbb{P}\left(\left(\bigcap_{\substack{j=1 \\ j \neq i}}^{f^g} E^{i, j}\right) \cap E^{i, l^g}\right)\right].$$

THEOREM 2 (HIT PROBABILITY FOR CLIENTS). *Consider a cache of size b that follows the Least Recently Used (LRU) eviction policy and serves a grouped client request pattern (see Section 2.2 for details). Under Approximations 1 and 2, the hit probability for a data object d requested by leader $l^g$ of group g is given by*

$$p^L(d, g) = 1 - \left[1 - p(d, t_b^*)\right] \left[\mathbb{P}\left(\bigcap_{i=1}^{f^g} E^{l^g, i}\right)\right]. \tag{10}$$

*Similarly, under Approximations 1 and 2, the hit probability for a data object d requested by i-th follower of group g is*

$$p^F(d, g; i) = 1 - \left[1 - p(d, t_b^*)\right] \left[\mathbb{P}\left(\left(\bigcap_{\substack{j=1 \\ j \neq i}}^{f^g} E^{i, j}\right) \cap E^{i, l^g}\right)\right]. \tag{11}$$

PROOF. To calculate the hit probability for a data object $d$ requested by the leader $l^g$ of group $g$, assume without loss of generality that this request occurs at time 0. The leader's hit probability depends on two factors: first, the randomly distributed request from $l^g$'s followers around time 0: $l^g$ can get a hit if at least one follower in group $g$ requests data object $d$ within $[-t_b^*, 0]$ since the followers can request for a data object before their leader does (Section 2.2 for details), second, requests other than the randomly distributed request from $l^g$'s followers around time 0: by Slivnyak's theorem [2], the remaining requests still follow a MPPP and thus $l^g$ can get a hit if at least one client, from any group, requests data object $d$ within the interval $[-t_b^*, 0]$ which is given by the earlier lemma under Approximations 1 and 2.

Using these two conditions, the leader's hit probability can be expressed as:

$$p^L(d, g) = 1 - \left[\mathbb{P}(\text{No follower in group } g \text{ requests } d \text{ in } [-t_b^*, 0])\right] \left[p(d, t_b^*)\right]. \tag{12}$$

From our earlier definition of event $E^{lg,i}$, we know:

$$\mathbb{P}(\text{No follower in group } g \text{ requests } d \text{ in } [-t_b^*, 0]) = \mathbb{P}\left(\bigcap_{i=1}^{f^g} E^{lg,i}\right). \tag{13}$$

Now, consider the hit probability for a data object $d$ requested by the $i$-th follower of group $g$. Again, assume without loss of generality that this request occurs at time 0. As before the follower's hit probability depends on two considerations: first, requests from clients other than $i$-th follower in group $g$: $i$-th follower can get a hit if at least one other client in group $g$ requests data object $d$ within $[-t_b^*, 0]$, second, requests other than the requests already considered: again using Slivnyak's theorem [2], the remaining requests still follow a MPPP and thus $i$-th follower can get a hit if at least one client (from any group) requests the data object $d$ within the interval $[-t_b^*, 0]$ which is given by the earlier lemma under Approximations 1 and 2.

Using these two conditions, we express the hit probability for follower $i$ as:

$$p^F(d, g; i) = 1 - \left[\mathbb{P}(\text{No client except } i\text{-th follower in group } g \text{ requests } d \text{ in } [-t_b^*, 0])\right]\left[1 - p(d, t_b^*)\right]. \tag{14}$$

From our earlier definition of events, we obtain:

$$\mathbb{P}(\text{No client except } i\text{-th follower in group } g \text{ requests } d \text{ in } [-t_b^*, 0]) = \mathbb{P}\left(\left(\bigcap_{\substack{j=1 \\ j \neq i}}^{f^g} E^{i,j}\right) \cap E^{i,lg}\right). \tag{15}$$

$\square$

In the coming section, we will use this theorem to understand the impact of client group structures.

REMARK 3 (CALCULATIONS FOR DIFFERENT DISTRIBUTION(S) OF $\Delta_i^g$). *We derive the following expressions for hit probabilities based on different distribution(s) of $\Delta_i^g$:*

(1) **Structured follower requests:** *If $\Delta_i^g = i \cdot \delta$, where $\delta > 0$, we can compute the concerned probabilities as follows:*

$$p(d, t_b^*) = 1 - e^{-\left(\sum_{g \in \mathcal{G}^d} \lambda^g(d)\left(t_b^* + f^g \cdot \min(\delta, t_b^*)\right)\right)}, \tag{16}$$

$$p^L(d, g) = p(d, t_b^*), \tag{17}$$

$$p^F(d, g; i) = \begin{cases} 1, & \text{if } \delta < t_b^*, \\ p(d, t_b^*), & \text{if } \delta \geq t_b^*. \end{cases} \tag{18}$$

## 2.4 Caching Policies

We will consider the set $\Pi$ of stationary caching policies including both online and offline policies which possibly adapt the cached content based on incoming requests, and may have knowledge about future requests or request rates. For a given vector of request rates $\Lambda$ and policy $\pi \in \Pi$, we define $\mathbf{h}_{\Lambda,\pi} = (h_{\Lambda,\pi}^{g,c}(d) : g \in \mathcal{G}, c \in C^g, d \in \mathcal{D}^g)$, where $h_{\Lambda,\pi}^{g,c}(d)$ denotes the long-term fraction of requests for data object $d$ from client $c$ of group $g$ that result in a cache hit.

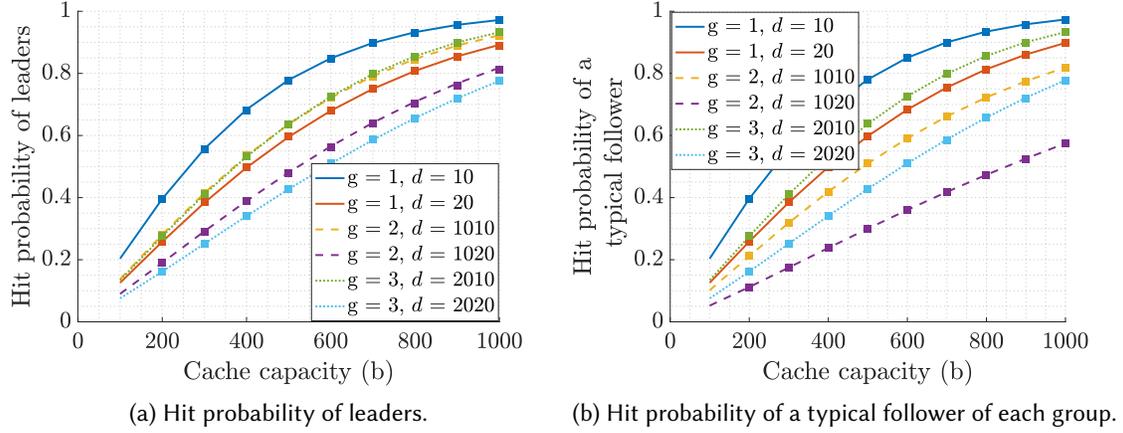

Fig. 2. Hit probability against cache capacity for different clients and data objects under LRU.

### 2.5 Performance Metric

We evaluate the performance of a caching policy based on the overall cache hit ratio. For a given request rate vector $\Lambda$ and policy $\pi$, the cache hit ratio is defined as:

$$H_{\Lambda,\pi} = \sum_{g \in \mathcal{G}} \sum_{d \in \mathcal{D}^g} \lambda^g(d) h_{\Lambda,\pi}^{g,l^g}(d) + \sum_{g \in \mathcal{G}} \sum_{i=1}^{f^g} \sum_{d \in \mathcal{D}^g} \lambda^g(d) h_{\Lambda,\pi}^{g,i}(d). \tag{19}$$

### 2.6 An optimal (offline) static caching policy

An optimal static caching policy is one that decides which fixed set of data objects to retain in the cache to maximize cache hit ratio given the request rate vector $\Lambda$. Let $\mathbf{x} = (x(d) : d \in \mathcal{D})$, where $x(d)$ is a binary variable indicating whether data object $d$ is cached. We formulate the following optimization problem to maximize the cache hit ratio:

$$\max_{\mathbf{x}} \quad \sum_{g \in \mathcal{G}} \sum_{d \in \mathcal{D}^g} \lambda^g(d) x(d)(1 + f^g) \tag{20a}$$

$$\text{s.t.} \quad \sum_{d \in \mathcal{D}} x(d) s(d) \leq b, \tag{20b}$$

$$x(d) \in \{0, 1\}, \forall\, d \in \mathcal{D} \tag{20c}$$

where constraint Eq. 20b ensures that the cache size does not exceed the capacity $b$.

### 2.7 Simulation results

In this subsection, we show the accuracy of the working set approximation for grouped client request pattern under the LRU caching policy and perform a comparative evaluation of different caching policies.

*2.7.1 How accurate is the working set approximation for grouped client request pattern under LRU?.*
**Setup.** We consider a caching system with three groups ($G = 3$), where each group can request data objects from a distinct set. The data object sets for the groups are defined as follows:

$$\mathcal{D}^1 = \{1, 2, \ldots, 1000\}, \quad \mathcal{D}^2 = \{1001, 1002, \ldots, 2000\}, \quad \mathcal{D}^3 = \{2001, 2002, \ldots, 3000\}.$$

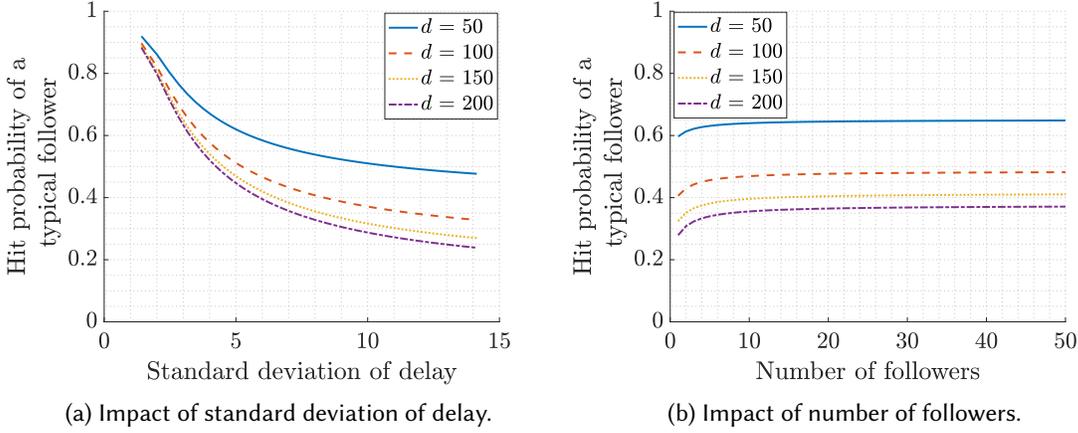

Fig. 3. Impact of client group structure on hit probability for follower requests.

Each group consists of one leader and several followers. The number of followers in each group is:
$$f^1 = 6, \quad f^2 = 4, \quad f^3 = 3.$$
The delay between the $i$-th follower (for all $i \in \{1, \cdots, f^g\}$) and the leader of group $g$ is modeled as an independent and identically distributed (i.i.d.) random samples from a uniform distribution:
$$\text{Group 1: } \Delta_i^1 \sim U[-10, 20], \quad \text{Group 2: } \Delta_i^2 \sim U[15, 30], \quad \text{Group 3: } \Delta_i^3 \sim U[-5, 40].$$
The request probability $p^g(d)$ for an object $d$ in group $g$ follows a Zipf distribution with parameter 1, identical across all groups. Requests from group leaders arrive according to a Poisson process, with the following rates:
$$\lambda^1 = 10, \quad \lambda^2 = 8, \quad \lambda^3 = 12.$$
The size of objects is defined so that even-numbered objects have a size of 2, while odd-numbered objects have a size of 5 for all groups.

**Results Discussion.** Fig. 2 shows the probability of hit for various objects requested by clients (leaders and followers) in different groups. Since follower delays in each group are i.i.d., the hit probability for all followers in a group is identical. The squares in the figure represent simulation results for the LRU policy, obtained from long simulation runs to ensure reliable accuracy. The lines are derived from the working set approximation for LRU. As shown, the approximation aligns almost perfectly with the simulation results across all clients, demonstrating its high accuracy for practical applications.

*2.7.2 Impact of client group structure on hit probability.* In this subsection, we use the working set approximation to analyze how the temporal overlap in follower requests and the number of followers affect the hit probability for follower requests. We focus on the following simulation setup:

**Setup.** We consider a caching system with a single group of clients requesting objects from the set $\mathcal{D}^1 = \{1, 2, \ldots, 5000\}$. The group has $f^1$ followers, and the delay between the $i$-th follower and the group leader is modeled as a uniform random variable $\Delta_i^1 \sim U[\alpha, \beta]$. The probability of a request $p^1(d)$ for a data object $d$ follows a Zipf distribution with parameter 1. The group leader generates requests according to a Poisson process with rate $\lambda^1 = 20$. Object sizes are defined such that even-numbered objects have a size of 2, while odd-numbered objects have a size of 5.

12et al.

**Impact of standard deviation of delay.** We fix the mean delay for follower requests, i.e., 0.5 $(\alpha + \beta) = 30$ and vary the standard deviation of the delay, $(\beta - \alpha)/\sqrt{12}$. The number of followers is set to $f^1 = 4$. Fig. 3a shows the hit probability of a typical follower for different data objects as a function of standard deviation of delay. Since follower delays are independent and identically distributed, all followers have the same hit probability. As the delay variance increases (reducing the temporal locality in follower requests), the hit probability for each data object decreases. This demonstrates how temporal locality can be quantitatively linked to cache performance under the LRU policy.

**Impact of number of followers.** Here, we fix $\alpha = 0$ and $\beta = 60$, and vary the number of followers. Fig. 3b depicts the hit probability for different data objects as a function of the number of followers. As the number of followers increases, the likelihood of requests for the same data object with short delays between them increases. Consequently, the hit probability for each data object increases.

## 3 CACHING BASED ON INFERRED CAUSAL RELATIONS

In the previous section, we analyzed the performance of the LRU under the grouped client request model. As we will observe in the simulation results, LRU performs suboptimally for small cache capacities compared to alternative caching strategies. To address this limitation, we propose a caching policy that infers temporal causal relationships between client requests and prioritizes eviction based on two key factors: (i) the frequency with which a client's request is followed by requests from other clients and (ii) the recency of requests. By leveraging these inferred dependencies, the proposed policy improves cache performance across different cache capacities.

Our focus is on structured following, such as the interactions between teachers (leaders) and students (followers) navigating a VR environment. However, the grouping of clients is not known a priori and must be dynamically inferred and tracked by our caching policy. In the following section, we introduce the necessary notation and framework to formalize this approach.

### 3.1 Notation

Recall that the set of clients is denoted by $C = \{1, 2, \ldots, C\}$, where $C = |C|$ represents the total number of clients. Let $a = ((a_m, d_m, c_m) : m \in \mathbb{Z}^+)$ represent a sequence of requests ordered in time, where $a_m$ is the arrival time of the $m$-th request, $d_m$ is the data object requested at that time, and $c_m$ is the client who made the $m$-th request. We denote the sequence of arrival requests generated by client $c$ as $a^c = ((a_n^c, d_n^c) : n \in \mathbb{Z}^+)$, where $a_n^c$ is the arrival time of the $n$-th request made by client $c$, and $d_n^c$ is the data object requested at that time. We let $M(t)$ denote the set of data objects present in the cache memory at time $t$.

To identify the last client who requested a specific data object $d$ before time $t$, we define the function:

$$c(d, t) = \begin{cases} \operatorname{argmax}_c \{a_n \cdot \mathbf{1}(d_n = d) \cdot \mathbf{1}(a_n < t) : c \in C, n \in \mathbb{Z}^+\} & \text{if such a client exists,} \\ -1 & \text{otherwise,} \end{cases} \quad (21)$$

where $\mathbf{1}(\cdot)$ is the indicator function. If no such client exists, the function returns -1.

Next, we define $n^c(t)$ as the index of the last request made by client $c$ at or before time $t$:

$$n^c(t) = \operatorname*{argmax}_n \left\{ a_n^c \leq t : n \in \mathbb{Z}^+ \right\}. \quad (22)$$

Definition 2 (Following event). *We define a following event, where say client $c_2$ follows $c_1$ on a cache hit if: (1) $c_2$ requests a data object for which it experiences a cache hit and (2) $c_1$ is the client that most recently requested the same data object.*

Formally, suppose that client $c_2$'s $n$-th request for data object $d_n^{c_2}$ at time $a_n^{c_2}$ results in a cache hit, i.e., $d_n^{c_2} \in M(a_n^{c_2})$, and that client $c_1$ was the last client to request $d_n^{c_2}$ before time $a_n^{c_2}$, i.e., $c(d_n^{c_2}, a_n^{c_2}) = c_1$. In this case, we say that client $c_2$ followed client $c_1$.

## 3.2 Caching policy

*3.2.1 Least Following and Recently Used (LFRU (w)).* The LFRU policy can improve on traditional LRU caching policy by considering inferred temporal relationships between clients' requests. This policy manages cache evictions by looking at both the recency of a request and how often other clients have followed the client who made the request. The goal is to keep data objects that were recently accessed and are more likely to be accessed again soon, based on these past patterns of client requests.

To quantify these temporal relationships, we examine the last $w$ requests made by each client, where $w$ is a parameter of the policy. Specifically, we construct a matrix $F(t)$ at time $t$ of size $C \times C$, where the entry in the $c_1$-th row and $c_2$-th column represents the number of times in the last $w$ requests of client $c_2$, that client $c_2$ followed client $c_1$. More formally, we define the $(c_1, c_2)$ element of the matrix as

$$F^{c_1,c_2}(t) = \begin{cases} \sum_{n=n^{c_2}(t)-w}^{n^{c_2}(t)} \mathbf{1}(d_n^{c_2} \in M(a_n^{c_2})) \mathbf{1}(c(d_n^{c_2}, a_n^{c_2}) = c_1) & c_1 \neq c_2 \\ 0 & \text{otherwise} \end{cases} \quad (23)$$

This matrix helps determine which clients' requests should be kept in the cache longer, based on the following event count for clients in the past.

**Description of the Policy:** When a new request $(a_n, d_n, c_n)$ arrives, we first check whether it results in a cache hit or a cache miss. If it is a cache hit, the requested object is moved to the most recent position in the access order to reflect its recency. If it is a cache miss, the object is added to the cache as the most recent entry, and evictions are performed as required.

To determine which object(s) to evict at time $a_n$, we first update the matrix $F(a_n)$, which captures the number of following events between pairs of clients. Next, using this matrix, we calculate the maximum number of times that other clients have followed each client's requests. Finally, objects associated with clients having the fewest following events are evicted first. If multiple objects correspond to clients with the same following event count, least recently accessed object is evicted. Note if no following events have been seen then the caching policy corresponds to LRU.

Remark 4 (Advantages of a Request-Based Window Over a Time-Based Window). *Using a fixed request-based window of the last $w$ requests per client, rather than a time-based window, provides a fair and consistent method for comparing following events across different clients. The key advantages of this approach are as follows:*

- **Fair comparison across different request rates:** *A request-based window ensures that following events are counted in a comparable manner for all clients. In contrast, a time-based window may count more events for clients with higher request rates, overestimating their influence on caching decisions.*
- **Adaptability to traffic variations:** *Since a request-based window directly tracks request behavior, it remains unaffected by fluctuations in request timing, making the caching policy more robust to dynamic traffic patterns.*

- ***Consistent eviction decisions:*** *By maintaining statistics based on a fixed number of past requests, a request-based window leads to stable and predictable cache eviction rules. In contrast, a time-based window varies with traffic conditions, potentially resulting in unpredictable cache behavior.*

*3.2.2 Least Following and Recently Used with Smoothing (LFRUS (w, γ)).* In the case that requests patterns are changing frequently, we consider a variant of LFRU policy, LFRUS, that adapts and assigns different weights to following events based on their recency. This policy is a combination of exponential averaging and sliding window. The key difference is in the calculation of $F^{c_1,c_2}(t)$, which is calculated as follows:

$$F^{c_1,c_2}(t) = \begin{cases} \left\lfloor \sum_{n=n^{c_2}(t)-w}^{n^{c_2}(t)} \gamma^{(n^{c_2}(t)-n)} \mathbf{1}(d_n^{c_2} \in M(a_n^{c_2})) \mathbf{1}(c(d_n^{c_2}, a_n^{c_2}) = c_1) \right\rfloor & c_1 \neq c_2 \\ 0 & \text{otherwise.} \end{cases} \quad (24)$$

Here, $\gamma$ is a parameter that determines the weight assigned to each following event, with more recent events given higher importance. The notation $\lfloor x \rfloor$ represents the floor of $x$. This policy has two parameters - $w, \gamma$.

## 4 SIMULATION RESULTS

In this section, we evaluate the proposed caching policy and compare its performance against standard caching policies across different request traces. These traces range from grouped client request model to client requests for objects within a VR environment.

For the VR-based request traces, clients move in groups and request objects that fall within their visibility range. We consider two approaches to generating these requests. First, we simulate client movement within a toroidal space where data objects are randomly distributed, and all objects have equal size. Second, we use an actual VR environment where the size of data objects depends on factors such as the number of vertices, edges, textures, and the distance of the client from the object.

The primary evaluation metric used in our analysis is the *cache hit ratio*, which measures the fraction of client requests that are successfully served from the cache.

**Caching policies.** We consider three additional online caching policies for evaluation.

- **Least Frequently Used (LFU):** LFU prioritizes caching data objects that have been accessed most frequently.
- **Belady:** Belady's algorithm evicts the data object that will be requested furthest in the future, making it an optimal but non-causal policy, as it requires full knowledge of the request sequence. We evaluate this policy for traces where all data objects have the same size.
- **Sieve** [21]: Sieve is a caching policy that retains recently accessed objects while efficiently managing evictions using a single queue and a "hand" pointer. Each object is marked as either visited or non-visited, and the least recently visited object with its visited bit unset is evicted. Similar to Belady, we evaluate this policy only for traces where all data objects have the same size.

### 4.1 Simulations for the grouped client request model

**Setup.** We consider a caching system with three groups ($G = 3$), where each group can request data objects from a distinct set. The sets of objects for the groups are defined as follows:

$$\mathcal{D}^1 = \{1, 2, \ldots, 1000\}, \quad \mathcal{D}^2 = \{1001, 1002, \ldots, 2000\}, \quad \mathcal{D}^3 = \{2001, 2002, \ldots, 3000\}.$$

Each group consists of one leader and several followers. The number of followers in each group is:

$$f^1 = 8, \quad f^2 = 6, \quad f^3 = 4.$$

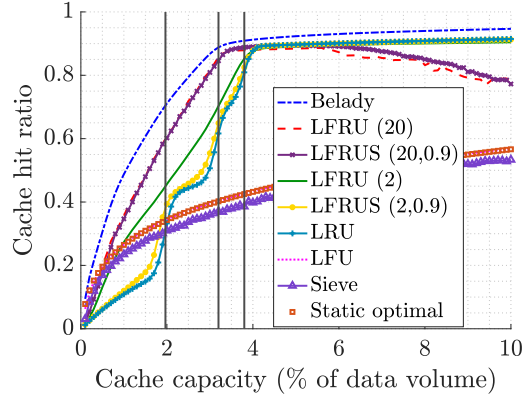

Fig. 4. Cache hit ratio against cache capacity defined as percentage of trace foot print (the number of unique objects in the trace) under different caching policies.

The delay between the $i$-th follower (for all $i \in \{1, \cdots, f^g\}$) and the leader of group $g$ is modeled as constant and ordered, with the following values:

$$\text{Group 1: } \Delta_i^1 = 10i, \quad \text{Group 2: } \Delta_i^2 = 20i, \quad \text{Group 3: } \Delta_i^3 = 30i.$$

The probability of a request $p^g(d)$ for an object $d$ in group $g$ follows a Zipf distribution with parameters 0.8, 0.85, and 0.9 for Groups 1, 2, and 3, respectively.

Requests from group leaders arrive according to a Poisson process, with the following rates:

$$\lambda^1 = 10, \quad \lambda^2 = 15, \quad \lambda^3 = 20.$$

The size of all data objects is 1.

**Results Discussion.** Fig. 4 presents the cache hit ratio for different caching policies across various cache capacities, expressed as a percentage of the total data volume (computed as the number of data objects multiplied by their average size). The cache capacity ranges from small (0.1%) to large (10%). As expected, Belady's policy, which has full knowledge of future requests, performs the best.

For larger cache capacities, LRU and LFRU/LFRUS with $w = 2$ outperform other online policies such as LFU, LFRU/LFRUS with $w = 20$, and Sieve, as well as the offline Static Optimal policy. This is because, with a larger cache, objects remain in cache for extended periods. Consequently, when a leader requests a data object, all its followers are more likely to experience a cache hit for the same object under LRU. However, the performance of LFRU/LFRUS with $w = 20$ degrades due to incorrect inferences of temporal causal relationships between clients. This issue arises for two primary reasons, first, leader requests are independent of past requests and follow a Zipf distribution, leading to frequent requests for popular objects, and second, as cache capacity increases, objects remain in the cache for longer, increasing the likelihood that a leader repeats a request for an object previously requested by another client. As a result, LFRU incorrectly infers that the leader follows that client. In contrast, with a smaller window size ($w = 2$), such incorrect inferences are quickly forgotten, minimizing their impact on caching decisions.

For small cache capacities, prioritizing data objects based on client-following behavior and request recency becomes crucial. In this case, LFRU/LFRUS effectively detects a limited number of following patterns and uses them to make more informed eviction decisions. This results in a significant performance gap between LFRU/LFRUS with $w = 20$ and $w = 2$ compared to LRU. A larger window size further widens this gap, as inferred following patterns persist longer, affecting eviction choices.

Under the grouped client request model, LRU performs well when the characteristic time, $t_b^*$—the time before a newly introduced object is evicted if there are no additional requests for this object—is longer than the delay between follower requests. In such cases, every follower request results in a cache hit, regardless of the requested object. To illustrate this effect, vertical lines in Fig. 4 mark cache capacities of 2%, 3.2%, and 3.8% of the total data volume. These correspond to cache sizes where the characteristic time matches follower delays in Groups 1, 2, and 3, with delays of 10, 20, and 30 units, respectively. As expected, around these points, the LRU hit rate exhibits a sharp increase.

In contrast, under LFU, followers experience a cache hit only when they request the most frequently accessed data objects that remain in the cache. Additionally, due to the high number of distinct requests, LFU performs similarly to the Static Optimal policy. Similarly, Sieve prioritizes the retention of frequently and recently requested objects. However, as demonstrated in our results, these strategies are suboptimal in this setting because follower requests closely track leader requests, even for objects that are infrequently accessed overall.

While in this section, we examined client request patterns based on the Grouped Client Request Model, this model assumes that a leader's requests are independent of past requests. However, in practical scenarios, client requests are inherently influenced by movement and past interactions, particularly in environments such as VR. To better capture these real-world dependencies, in the next section, we shift our focus to a synthetic VR environment, where client requests arise as users navigate a shared space.

### 4.2 Simulations for synthetic request traces for client motion in a Toroid

*4.2.1 Simulation environment for generating cache request traces.* To analyze spatio-temporal patterns in data object requests, we generate cache request traces in a controlled simulation environment where clients move as part of distinct groups. The simulation takes place in a 3D toroidal cube, with each side measuring 1000 units. This space contains 4000 data objects, which are randomly distributed throughout the volume. The size of each data object is equal; we revisit this in the next section.

To model group motion dynamics, each group consists of a leader client and multiple follower clients. The total number of groups, the designated leaders, and the overall number of clients remain fixed throughout the simulation. Leaders follow unique motion paths, while followers trail behind with a fixed delay. Group composition can change over time, with followers either reordering themselves within the group or switching to a different leader.

The simulation operates in discrete time slots, where each time slot represents one unit of simulated time. During each time slot:

(1) Leaders update their positions based on a constant speed of 25 units per time slot and their current direction. Every 10 time slots, each leader selects a new random direction. They move continuously through the toroidal space, reappearing on the opposite side when crossing a boundary.
(2) Followers update their positions by following the path of their assigned leader, but with a delay of a specified number of time slots. For example, if a follower has a delay of $\tau$ time slots, its position in the current time slot matches the leader's position from $\tau$ time slots earlier. This delay models how group members in the real world follow a leader with some lag.
(3) Each client, including both leaders and followers, requests data objects located within a 360-degree field of view with a radius of 50 units.

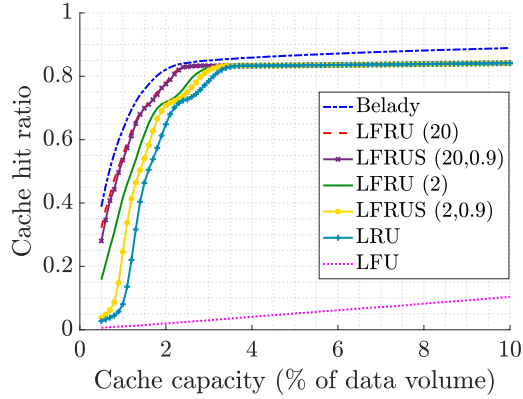

Fig. 5. Cache hit ratio against cache capacity under different caching policies for Trace 1.

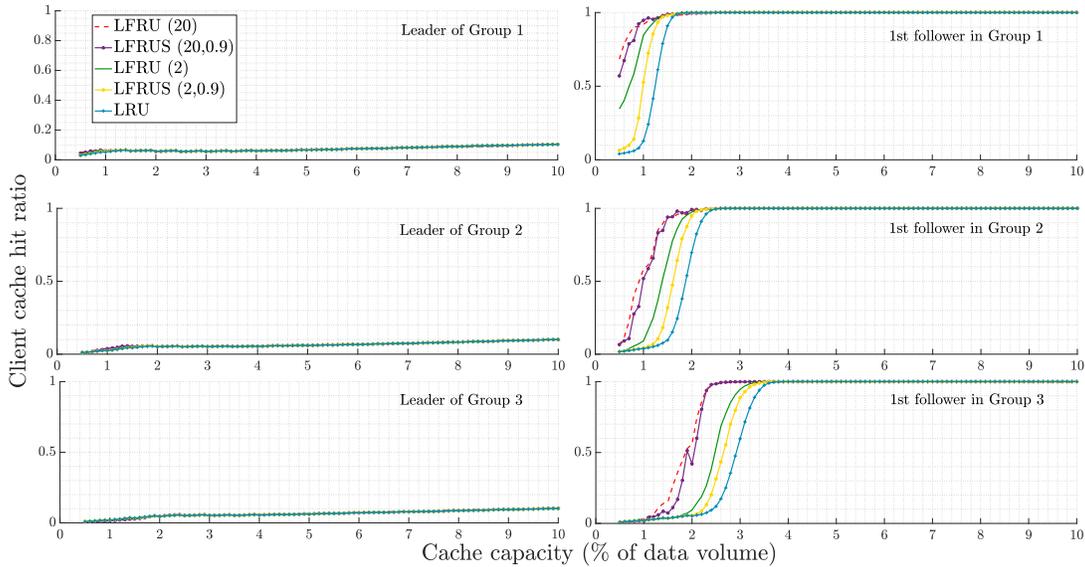

Fig. 6. Cache hit ratio for different clients against cache capacity under different caching policies for Trace 1.

The simulation runs for 10 million time slots, enabling the capture of long-term spatio-temporal patterns in client movement and data access.

*4.2.2 Trace 1: Static following.* **Setup.** In this trace, we simulate 3 groups comprising a total of 17 clients navigating the 3D toroidal space. Groups 1, 2, and 3 consist of 8, 4, and 2 followers, respectively. Each group's leader is initialized at a random location and navigates according to the dynamics described above. The delay between the $i$-th follower and the leader varies by group. In Group 1, the $i$-th follower trails its leader with a delay of $4i$ time slots, e.g., Follower 1 trails the leader by 4 time slots, Follower 2 by 8 time slots, and so on. Similarly, in Group 2, the $i$-th follower trails its leader with a delay of $8i$ time slots, while in Group 3, the delay is $20i$ time slots. Each client maintains a local cache managed using the Least Recently Used (LRU) policy. The size of each local cache is set to 5% of $b$, where $b$ is the total cache capacity of the edge or cloud server. When a client makes a request, it first checks its local cache. If the requested item is not found (a cache miss), the client forwards the request to the main cache (edge or cloud server). The cache hit ratio is defined as the fraction of requests to the main cache (edge or cloud server) that result in a hit.

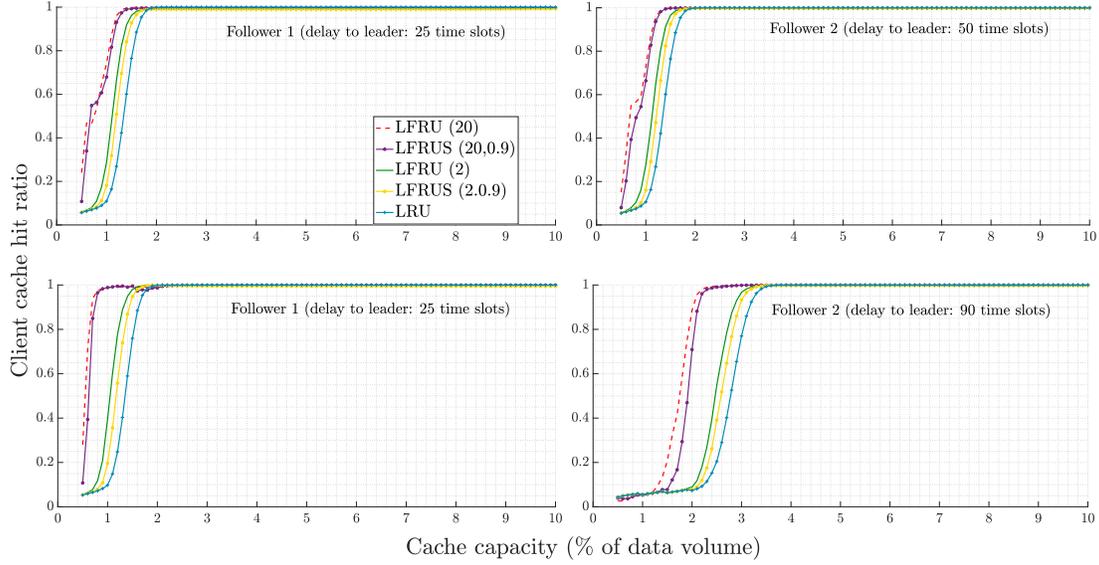

Fig. 7. Comparison of client cache hit ratios across different caching policies under two delay configurations: uniform delays between followers (row 1) and non-uniform delays (row 2).

**Results discussion.** Figure 5 shows the cache hit ratio for different caching policies at various cache capacities, expressed as a percentage of total data volume (computed as the number of data objects multiplied by their average size). The cache capacity ranges from small (0.1%) to large (10%). In Figure 6, we also show the client cache hit ratio for a selected set of caching policies. The client cache hit ratio is defined as the proportion of requests from a specific client that result in a cache hit, compared to the total number of requests. As expected, Belady's policy, which is based on the knowledge of future requests, performs the best. This is because it can retain objects that will be requested again soon.

Next, LFRU/LFRUS with $w = 20$, LFRU/LFRUS with $w = 2$, and LRU show similar performance when the cache capacity is large. This happens because, with a larger cache, objects stay in the cache for longer periods. As a result, once a leader requests a data object, all its followers also experience a cache hit for that object, as shown in Figure 6. Therefore, even though our policies account for following events to decide on evictions, there is enough cache space to keep the objects requested by each leader.

However, when the cache capacity is small and space is limited, it becomes important to decide which objects to keep based on client-following behaviors and the recency of requests. This creates a noticeable performance gap between LFRU/LFRUS with $w = 20$ and $w = 2$ as compared to LRU. As seen in Figure 6 for our LFRU-based policies, followers start to experience cache hits for their requests, which does not happen with LRU. With a larger window, this gap becomes even larger, as any detected following event stays in the window for a longer period of time. However, if older following events are not given equal weight, a performance gap appears, as seen in the difference between LFRU and LFRUS with the same window sizes.

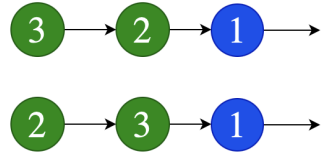

Fig. 8. After every p time slots, Client 2 and 3 (followers) swap their positions in following Client 1 (leader).

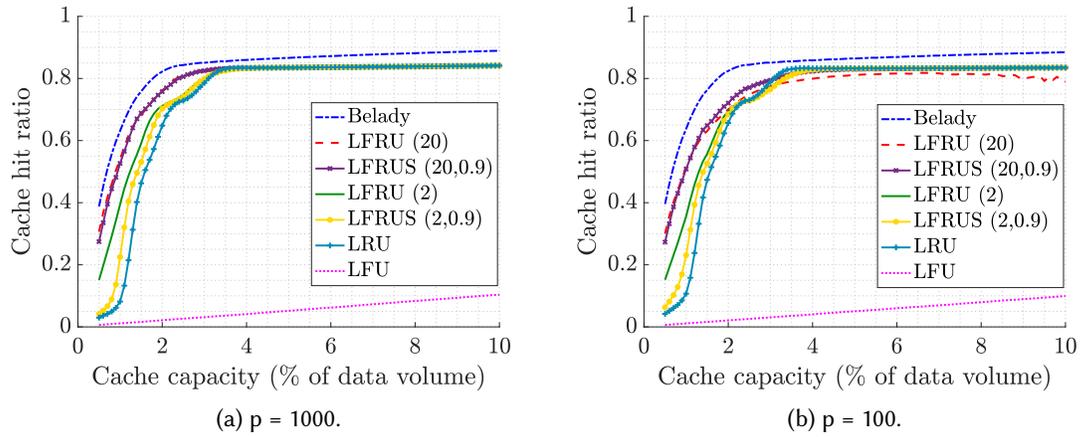

Fig. 9. Cache hit ratio against cache capacity under different caching policies for Trace 2.

We also observe performance jumps for LRU. These jumps occur when the delay between followers in a group becomes roughly equal to the time a typical object stays in the cache under LRU. We can see this in Figure 6, where the client cache hit ratio for followers in different groups shows jumps in performance under LRU at different cache capacities. These jumps follow the order of delays between followers in the groups, meaning that Group 1 followers, with lower delays, experience a higher cache hit ratio than Group 2 followers. Finally, the LFU policy, which prioritizes the most frequently accessed objects, performs poorly. This is because when a follower requests the same object as their leader, the object's access frequency increases, causing it to stay in the cache longer than needed. Instead, the policy should focus on keeping new data objects requested by group leaders, even if they are rarely accessed.

*4.2.3 Effect of different delays between followers for Static following.* **Setup.** In this simulation, we consider a group consisting of three clients: one leader and two followers. We examine two scenarios for the delay between followers - Case 1: The delay between followers is uniform. The $i$-th follower trails its leader with a delay of $25i$ time slots, Case 2: The delay between followers is non-uniform. Follower 1 trails its leader with a delay of 25 time slots, while Follower 2 trails its leader with a delay of 90 time slots.

**Results discussion.** Fig. 7 presents the client cache hit ratio for different caching policies at various cache capacities. In the first row, which corresponds to Case 1, both Follower 1 and Follower 2 exhibit similar performance. However, in the second row, which represents Case 2 with increased delay for Follower 2, we observe a decline in Follower 2's cache hit ratio compared to Case 1. This performance drop is consistent across all caching policies.

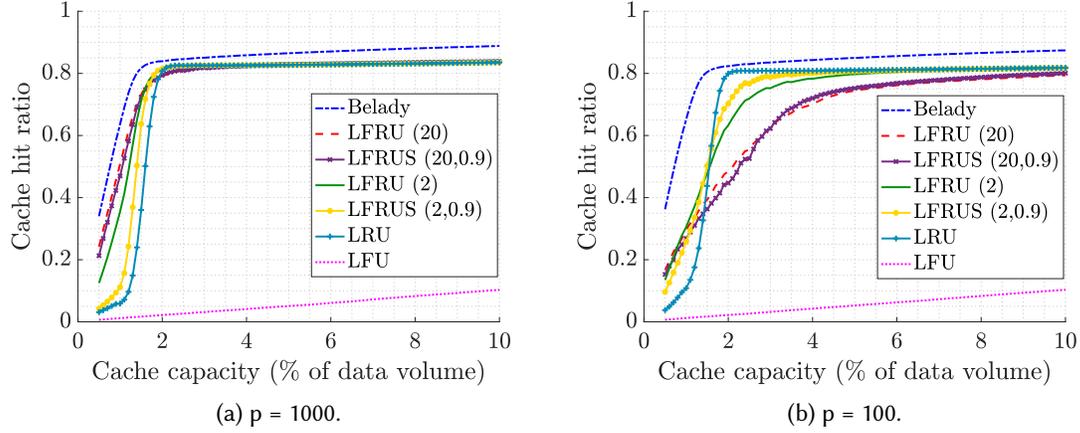

Fig. 10. Cache hit ratio against cache capacity under different caching policies for Trace 3.

*4.2.4 Trace 2: Periodic order shuffling.* **Setup.** In this trace, the group characteristics remain the same as in Trace 1. However, we introduce periodic shuffling of followers, denoted by a parameter $p$, which specifies the number of time slots in each period. At the start of a new period, follower $2i − 1$ and follower $2i$ in each group exchange their delays/positions relative to the leader, see Fig. 8. This value of the parameter $p$ allows us to evaluate the impact of varying the frequency of these swaps on the performance of our caching policies. As before, each client has a local cache managed using the Least Recently Used (LRU) policy. The size of each local cache is set to 5% of $b$, where $b$ represents the total cache capacity (of edge/cloud server).

**Results discussion.** Figures 9a and 9b show the cache hit ratio for various caching policies when the period $p$ is large and small, respectively, indicating how long a particular following behavior persists.

When $p$ is large, the proposed caching policies that utilize inferred causal relationships continue to perform well. However, there is a slight performance degradation (though not noticeable) compared to Trace 1.

When $p$ is small, the performance of LFRU with $w = 20$ decreases for larger cache sizes as compared to Trace 1 in the previous section. This happens because followers change their order more frequently, so the policy is biasing its decisions on incorrect inferences of following relationships. Additionally, the cache needs to flush out all data objects associated with clients that were previously perceived as being followed by others, but are no longer followed. This leads to a performance loss.

This issue can be mitigated by using a smaller window, as shown by LFRU with $w = 2$. A smaller window helps the policy adapt more quickly by forgetting past following events. Alternatively, assigning weights to following events ensures that only persistent following events influence the decision of which data objects to retain; see LFRUS variants of LFRU.

For small cache capacities, larger windows remain preferable because the likelihood of LFRU detecting all following patterns is lower. As a result, even when the order changes, there is no significant performance degradation.

*4.2.5 Trace 3: Periodic leader switching.* **Setup.** In this trace, the number of leaders remains 3, and the total number of clients remains 17. However, we introduce periodic leader changes for followers, controlled by a parameter $p$, which specifies the number of time slots in each period. At the start of a new period, each follower selects a new leader with predefined probabilities: Leader

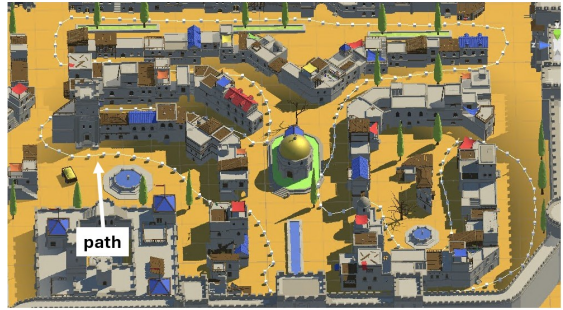

Fig. 11. The virtual city of the simulation.

1 is chosen with probability 0.5, Leader 2 with probability 0.3, and Leader 3 with probability 0.2. When a follower selects a new leader, its delay relative to this new leader is determined by the number of followers who have already selected this leader and the delay between consecutive followers, which we set to 5. Consequently, the $i$-th follower trails behind its newly selected leader by $5i$ time slots. As before, the parameter $p$ allows us to study the effect of varying the frequency of these leader switches on the performance of our caching policies. Lastly, each client has a local cache managed using the Least Recently Used (LRU) policy. The size of each local cache is set to 5% of $b$, where $b$ represents the total cache capacity (of edge/cloud server).

**Results discussion.** Figures 10a and 10b show the cache hit ratio for various caching policies when the period $p$ is large and small, respectively. The period $p$ indicates how long a group configuration and the order of followers in that group persist.

When duration is large, as shown in Fig. 9a, our caching policies that use inferred causal relationships continue to perform well.

When the duration is small, there is a significant performance gap between LFRU with $w = 20$ and LRU for medium to large cache sizes. This occurs because, in this case, followers not only change their order, but also switch leaders. With a larger window, it takes more time to forget and update the cache. This performance gap decreases when using a smaller window, such as LFRU with $w = 2$.

However, the best performance is achieved when we also apply smoothing along with a small window. This allows the policy to adapt quickly and giving more weightage to recent following events between clients. This approach also shows a significant performance improvement over LRU, especially for smaller cache capacities.

The synthetic VR environment in this section captures how client requests depend on movement, but does not account for certain real-world factors. In practice, objects in a VR scene may be occluded by other objects, making them temporarily inaccessible. Additionally, the size of an object in a request depends on its pixel size, texture, geometric complexity, and distance from the client. These factors affect how objects are rendered and influence caching decisions. In the next section, we refine our request model to incorporate these additional constraints, making it more representative of real-world VR environments.

### 4.3 Simulations for emulated VR request traces

*4.3.1 Simulation environment for generating cache request traces.* We generated cache request traces using a simulator that models a virtual desert city, measuring $376.29 \times 608.22$ m$^2$ and containing 1,176 objects (e.g., buildings, trees, fountains). In this simulation, a virtual reality (VR) client navigates the city from a height of 1 to 2 meters above the ground, which represents the typical height of a VR headset, whether the client is sitting or standing. Objects in the city may be occluded by buildings depending on the client's position, so clients can only request visible data.

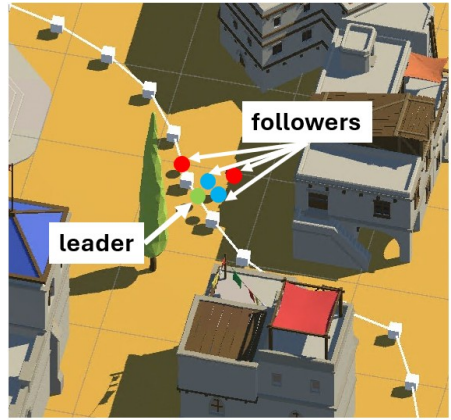

Fig. 12. Example relative positions of the leader and followers in unstructured follower requests.

To determine visibility, we dynamically compute the set of objects that are visible from all directions around the client's current position. This approach is preferred over relying on a fixed or random view direction, as it ensures that we send data for all possible directions rather than waiting for the client to specify their view. This method is particularly important because visibility computation is computationally expensive, and the client's viewpoint may change rapidly. Computing visibility only for the current view direction would be inefficient, especially when the client may quickly turn or move.

Clients navigate through alleys along pre-designated, looped paths, avoiding collisions with objects, as shown in Fig. 11. In our dataset, there are always three client groups, each consisting of five clients and each group follows a distinct path. Within each group, one client serves as the leader, while the remaining clients follow in the same direction. These paths are bidirectional and non-intersecting, and all clients move at a constant speed of 2 m/s.

The simulation operates in discrete time slots. In each slot, each client updates its position based on the elapsed time since the previous slot and recomputes the visibility of objects. A client initiates a request for an object if it becomes visible in the current time slot and was not requested in the previous slot. This request mechanism assumes that the client's local cache is large enough to store data retrieved in a given time slot for use in the subsequent slot.

In the simulation, data objects come in multiple versions, each varying in size. The version of a requested data object depends on the distance of the client from the object: If the client is within 10 m, it requests the highest quality version; if it is beyond 50 m, it requests lowest quality version; and for distances between 10 and 50 m, it requests middle quality version. The size of each version is influenced by factors such as pixel resolution, texture, and geometric detail. For example, the highest quality version of an object may be 1MB, the middle quality version 0.5MB, and the lowest quality version 0.1MB. Additionally, each request must be served with the exact version specified by the client; a request for one version cannot be fulfilled using another version of the same object.

*4.3.2 Trace 1: Unstructured follower requests.* **Setup.** In this trace, all clients within the same group are initialized near a common point on their designated path (Fig. 12). Each client starts at a random position within 4 meters of this point. Clients do not enter buildings during navigation.

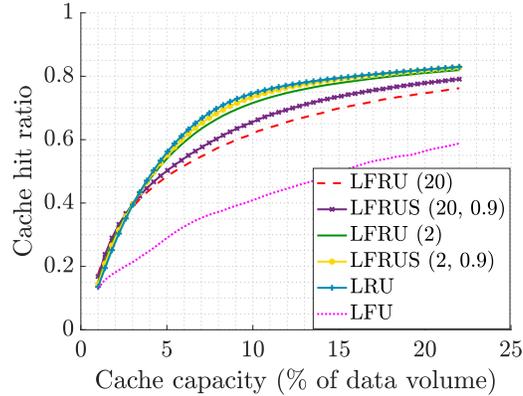

Fig. 13. Cache hit ratio against cache capacity under different caching policies for unstructured follower requests.

All followers move in the same direction as their group's leader, maintaining a constant delay of 0.5 seconds. Specifically, if the leader changes direction at a corner, all followers adjust their movement 0.5 seconds later. Each client maintains a local cache of previously requested objects managed using the LRU policy. The size of each local cache is set to 5% of $b$, where $b$ is the total cache capacity of the edge or cloud server. When a client makes a request, it first checks its local cache. If the requested item is not found (a cache miss), the client forwards the request to the main cache (edge or cloud server). We have considered a local cache for each client as described for all simulation results discussed hereafter.

**Results discussion.** Fig. 13 presents the cache hit ratio for different caching policies across various edge/cloud cache capacities, expressed as a percentage of the total data volume (computed as the number of data objects multiplied by their average size). The cache capacity varies from small (1%) to large (22%).

Our results show that LRU performs best for moderate to large cache capacities. This is because, in unstructured following, followers are distributed around the leader in a staggered pattern and do not consistently request the same set of data objects. Consequently, the number of consistent following events is lower, making inferred temporal relationships less effective for cache management. In such cases, eviction decisions based on recency outperform those based on inferred causal relationships. This is evident from the performance gap between LFRUS and LFRU for the same window size, where smoothing enables the caching policy to assign greater importance to recent following events. Additionally, using a smaller window size improves adaptability by allowing the policy to quickly discard outdated following patterns. For small cache capacities, LFRU/LFRUS performs better with larger window sizes. Rather than considering the recency of all client requests, these policies prioritize clients that are closer and exhibit more persistent following behavior, leading to an improved cache hit ratio.

*4.3.3 Trace 2: Static following (Structured follower requests).* **Setup.** In this trace, all followers move along the path by following their leader (Fig. 14). The leader is initialized at a random point on the path, and each follower is subsequently initialized with a constant delay of 2 seconds.

For example, if the leader is initialized at position $x$, the first follower starts at $x$ - 4 meters along the path (assuming movement in a positive direction). The second follower is initialized at $x$ - 8 meters, and so forth. Throughout the simulation, all clients maintain their initial relative distances from one another on the path.

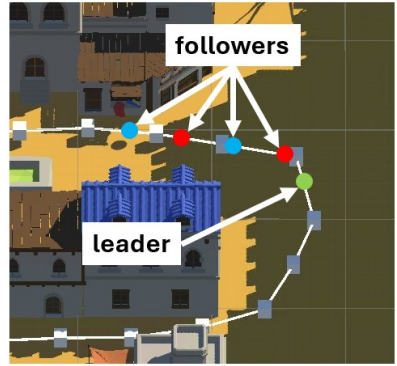

Fig. 14. Example relative positions of the leader and followers in structured follower requests.

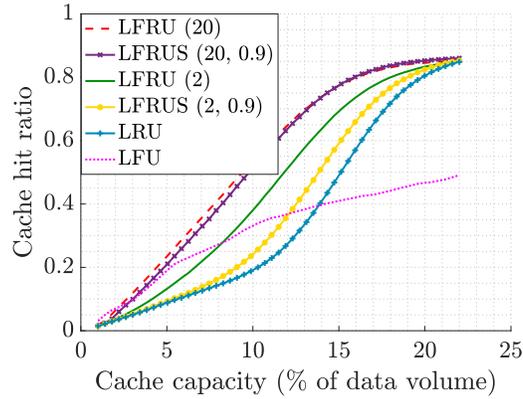

Fig. 15. Cache hit ratio against cache capacity under different caching policies for structured follower requests.

**Results discussion.** Fig. 15 presents the cache hit ratio for different caching policies across various cache capacities, ranging from small (1%) to large (22%).

Our results indicate that LFRU/LFRUS with $w = 20$ consistently outperforms all other caching policies across all cache capacities. This is because, in structured following, followers sequentially request data after their leader and tend to repeat the leader's requests. Once these following events are detected, they enhance cache eviction decisions by prioritizing the retention of objects requested by a client (excluding the last follower) within a group, as these objects are likely to be requested again by another follower in the same group.

The advantage of retaining the following events detected for longer is evident in the performance gap between LFRU with $w = 20$ and $w = 2$. A larger window size allows the policy to leverage persistent following patterns more effectively, leading to higher cache hit ratios. Furthermore, in this setting, smoothing is unnecessary, as following events remain stable over time.

*4.3.4 Trace 3: Periodic order shuffling.* **Setup.** In this trace, the setup remains the same as in Trace 2, except that followers periodically swap positions in pairs. Every $p$ seconds (30s or 150s), adjacent followers swap positions: the first follower swaps with the second, the third with the fourth, and so on.

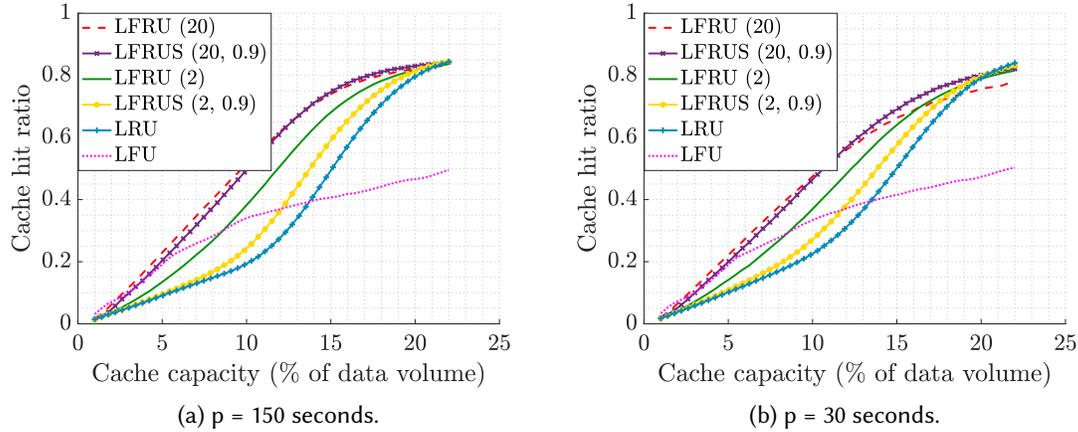

(a) p = 150 seconds.    (b) p = 30 seconds.

Fig. 16. Cache hit ratio against cache capacity under different caching policies for Periodic order shuffling.

The swap occurs gradually over a duration of 3 seconds, rather than instantaneously. During this process, clients continue their normal movement along the path, but their relative distances temporarily change as they transition to their new positions.

**Results discussion.** Figures 16a and 16b present the cache hit ratio for different caching policies when the period $p$ is large and small, respectively. The period $p$ represents the duration for which a particular following behavior persists.

When $p$ is large, caching policies that leverage inferred causal relationships maintain strong performance. However, a slight performance drop is observed compared to Fig. 15, though it remains negligible.

When $p$ is small, the performance of LFRU with $w = 20$ deteriorates for larger cache sizes compared to scenarios with a larger $p$. This decline occurs because followers change their order more frequently, leading the policy to make incorrect inferences about following relationships. Additionally, the cache must evict objects associated with clients who were previously identified as being followed but are no longer, resulting in performance degradation.

This issue can be mitigated by using a smaller window size, as demonstrated by LFRU with $w = 2$. A smaller window enables the policy to adapt more quickly by discarding outdated following events. Alternatively, assigning weights to following events ensures that only persistent relationships influence caching decisions, as seen in the LFRUS variants of LFRU.

For small cache capacities, larger window sizes remain beneficial, as LFRU is less likely to capture all following patterns. Consequently, even when follower order changes, the performance does not degrade significantly.

## 5 CONCLUSION

In this paper, we introduced a model for capturing correlated client request patterns. We then showed that correlations in client requests influence the characteristic time of an LRU-managed cache, leading to improved cache hit ratios compared to LFU as cache size increases. To further leverage these correlations, we proposed LFRU, an adaptive caching policy that dynamically infers and exploits causal relationships between client requests. By incorporating both the recency of requests and the frequency with which clients follow others' requests, LFRU enhances its eviction decisions. Our empirical evaluations on synthetic and VR-based datasets demonstrated that LFRU achieves up to 2.9× and 1.9× improvements in cache hit ratio over LRU and LFU, respectively, under structured following. These findings underscore the importance of incorporating request correlations in caching strategies and provide a foundation for the development of more adaptive and efficient caching mechanisms in future networked systems.

# APPENDIX